\documentstyle[preprint,aps,epsfig]{revtex}

\tightenlines

\begin{document}
\draft
\title{Generating vortex rings in Bose-Einstein condensates
in the line-source approximation}
\author{M. Guilleumas$^1$ , D.M. Jezek$^2$, R. Mayol$^1$,
M. Pi$^1$, M. Barranco$^1$}
\address{$^1$Departament d'Estructura i Constituents de la Mat\`eria,
Facultat de F\'{\i}sica, \\
Universitat de Barcelona, E-08028 Barcelona, Spain}
\address{$^2$Departamento de F\'{\i}sica, Facultad de Ciencias Exactas
y Naturales, \\
Universidad de Buenos Aires, RA-1428 Buenos Aires, and \\
Consejo Nacional de Investigaciones Cient\'{\i}ficas y T\'ecnicas,
Argentina}

\date{\today}

\maketitle

\begin{abstract}

We present a numerical method for generating vortex rings
in Bose-Einstein condensates confined in axially symmetric traps.
The vortex ring is generated using the line-source approximation
for the vorticity, i.e., the rotational of the superfluid velocity
field is different from zero only on a circumference of given radius
located on a plane perpendicular to the symmetry axis and
coaxial with it.
The particle density is obtained by solving a modified Gross-Pitaevskii
equation that incorporates the effect of the velocity field.
We discuss the appearance of density profiles, the vortex core
structure and the vortex nucleation energy, i.e., the energy difference
between vortical and ground-state configurations. This is used to
present a qualitative description of the vortex dynamics.

\end{abstract}


\narrowtext
\section{Introduction}

Since 1999, when vortex lines in a trapped Bose-Einstein condensate (BEC)
were first experimentally obtained \cite{ma99},
their study has received great  experimental  and theoretical
interest as it constitutes a clear signature of superfluidity effects
in these confined systems.
Some remarkable experimental achievements are, among others,
the study of the dynamics of single vortex lines
\cite{an00} and the formation of small \cite{da} and large \cite{ke}
vortex arrays. A review of the research done in this field
is presented in Ref. \cite{fe01}.
It is worth to note that the experimental
situation in BEC regarding the formation and detection of vortices
is at variance with that in helium II droplets. For
these drops, a  paradigm of superfluid finite systems, the problem
of nucleating vortices and their stability has only been
addressed very recently from the theoretical point of view (see Refs.
\cite{Bau95,Dal00,May01}), and their experimental detection is still an
open question.

Vortex rings  are vortices whose core is a closed loop with quantized
circulation around it \cite{Donnelly}. They are
complex topological structures that have attracted and will
continue to attract some experimental and theoretical interest.
Based on numerical simulations,
different methods have been proposed to generate vortex
rings in BEC. As in bulk liquid helium, vortex rings may
be produced introducing an impurity in the condensate
with a definite velocity, whose displacement causes
a vortex ring\cite{ja991}. Another mechanism\cite{fe00}
consists in using
dynamical instabilities in the condensate to cause dark solitons to
decay into vortex rings.  A well controlled method to produce vortex
rings by electromagnetically induced atomic transitions in
two-component condensates has also been put forward\cite{ru01}.
The method proposed in Ref. \cite{fe00} has been successfully applied
in Ref. \cite{an01} to generate  vortex rings experimentally.

Rather than proposing a method that could be implemented to
create a vortex ring experimentally, our aim here is to set up a
numerical method simple yet accurate enough to
generate quantized vortex rings with a definite radius $R$
in one-component condensates.  We restrict our study to vortex states
such that the divergence of their velocity is vanishingly small.
This assumption yields analytical expressions for the
velocity field, and provides a fair approximation to the
superfluid flow around the vortex core for rings in the bulk
of large condensates (Thomas-Fermi limit).
We have considered large condensates at zero temperature and therefore,
dissipation has not been taken into account. They are
axially symmetric about the $z$ axis, and have the $z=0$ plane as
symmetry plane, and may  host a vortex ring of radius $R$,
coaxial with the  symmetry axis of the trap, and placed on a plane at a
distance $Z\geq 0$ from the symmetry plane. Although generalization to
situations with more than one such vortex rings is straightforward,
we have not considered this possibility.

This paper is organized as follows. In Sect. II we present the
method used to generate vortices. The way to obtain the
velocity field of a vortex ring is described in Sect. III, and the
explicit expressions of the velocity components are given in an
Appendix. Sect. IV is devoted to the analysis of the particle density
profiles  and vortex nucleation  energies, which allows a
qualitative description of their dynamics.  We also present results
obtained from a completely different, more involved method we have
set up to generate ring  vortices that permits to test the approximation
of zero divergence of the velocity. Finally, a brief summary of the
results is presented in Sect. V.

\section{Energy functional}

We consider a weakly interacting Bose-condensed gas confined in a
harmonic trap $V_{\rm ext}({\bf r})$ at zero temperature.
In the Gross-Pitaevskii (GP) theory, the ground state (g.s.) energy of
the condensate is given by the functional \cite{da99}
\begin{equation}
E[\,\Psi\,] = \int d{\bf r} \left[\frac{\hbar^2}{2 m} |\nabla \Psi|^2 +
 V_{\rm ext}({\bf r}) |\Psi|^2 + \frac{g}{2} |\Psi|^4 \right] \,\,\, ,
\label{functional}
\end{equation}
where $\Psi({\bf r})$ is the condensate wave function.
The first term in Eq. (\ref{functional})
is the kinetic energy of the condensate, the second term
is the harmonic oscillator energy arising from the trapping potential,
and the third term is the mean-field interaction energy. The coupling
constant is $g=4 \pi \hbar^2 a /m$, where $a$ is the $s$-wave
scattering length, and $m$ is the atomic mass.
The number of atoms in the condensate is
$\int \! d{\bf r} |\Psi|^2= N$.
The g.s. wave function is determined by solving the GP equation
obtained minimizing the energy functional.

The wave function $\Psi$ can be written in terms of the particle density
$\rho({\bf r})=|\Psi({\bf r})|^2$ and phase $S({\bf r})$ as
\begin{equation}
\Psi({\bf r})=\sqrt{\rho({\bf r})} \, \exp[i S({\bf r})] \,,
\label{wf}
\end{equation}
and the superfluid velocity is given by ${\bf v}=(\hbar/m) {\bf \nabla}
S$. In this work we will use
the equivalent quantum hydrodynamic description
of the condensate in terms of the density and the
superfluid velocity\cite{da99}, since it
allows a straightforward generalization of the energy functional
Eq. (\ref{functional}) to include vortex states.
Using Eq. (\ref{wf}) it follows that
\begin{equation}
E[\,\Psi\,] =  E_0 [\,\rho\,] +
    E_{\rm kin}^{\bf v}[\,\rho,{\bf v}\,] \,.
\label{eqe}
\end{equation}
The first term is only density-dependent:
\begin{equation}
E_0 [\,\rho\,]=
\int \,d\,{\bf r} \left[ \frac{\hbar}{2 m} |\nabla \sqrt \rho|^2
  + V_{\rm ext}({\bf r}) \rho + \frac{g}{2} \rho^2 \right] \,\,\, .
\label{eqe0}
\end{equation}
The first term in $E_0[\rho]$ is the quantum kinetic energy.
The second term of the energy functional Eq. (\ref{eqe})
corresponds to the kinetic energy associated to the flow velocity
${\bf v}$, and is given by
\begin{equation}
E_{\rm kin}^{\bf v} [\,\rho,{\bf v}\,]  = \frac{1}{2} m
\int d {\bf r} \, \rho({\bf r}) \, {\bf v}^2({\bf r}) \,\,\, .
\label{eqkin}
\end{equation}
The g.s. wave function in the absence of vortices has a
spatially constant phase and therefore zero velocity.
It can be obtained by minimizing $E_0 [\,\rho\,]$.
We consider condensates with positive scattering length and
axially symmetric traps
$V_{\rm ext}({\bf r})=m [\omega_{\perp}^2 (x^2+y^2)+ \omega_z^2 z^2]/2$
$[=m (\omega_{\perp}^2 r^2+ \omega_z^2 z^2)/2$ in cylindrical $(r,z)$
coordinates], with different values of the
asymmetry parameter $\lambda=\omega_z/\omega_{\perp}$.
The trap harmonic frequency $\omega_{\perp}$ provides a
length scale for the system, $a_{\perp}=(\hbar/m \omega_{\perp})^{1/2}$.
We will use $a_{\perp}$, $\hbar \omega_{\perp}$, and $N/a_{\perp}^3$
as units of length, energy, and density, respectively.

If we consider large condensates in which the
Thomas-Fermi (TF) approximation holds\cite{da99},
the quantum kinetic energy can be neglected compared to the
interaction energy in Eq.~(\ref{eqe0}), and
the g.s. density of the condensate in the absence of vortices
is given by
$\rho_0(r,z)= \mu(1-r^2/R_{\rm TF}^2-z^2/Z_{\rm TF}^2)/g$
in the region where this expression is positive,
and zero elsewhere.
The TF  extents of the condensate in the radial and
axial directions are
$R_{\rm TF}=(2 \mu / m \omega_{\perp}^2)^{1/2}$ and
$Z_{\rm TF}=R_{\rm TF}/\lambda$.
The chemical potential $\mu$ is fixed by normalization
$\mu=\hbar \omega_{\perp}(15 \lambda a N/a_{\perp})^{2/5}/2$.
We recall that the validity of the Thomas-Fermi approximation
is guaranteed if $N a/a_\perp \gg 1$.

We turn now our attention to the case of condensates with vortex
states characterized by a given irrotational
velocity field associated with a non-vanishing quantized circulation.
The total energy of the system is given by
the energy functional Eq. (\ref{eqe}). If the
velocity field is known, for a given number of particles
the density profile can be obtained minimizing Eq.~(\ref{eqe}). This
yields the equation
\begin{equation}
\left( - { \hbar^2 \nabla^2 \over 2m} + V_{\rm ext}({\bf r})
  + g \mid \!\psi({\bf r}) \!\mid^2 + \frac{1}{2} m {\bf v}^2
   \right) \psi({\bf r}) = \mu \, \psi({\bf r}) \,\,\, ,
\label{GP}
\end{equation}
where $\psi=\sqrt \rho$ is the modulus of the complex
wave function Eq. (\ref{wf}). This is
the GP equation expressed in terms of the
hydrodynamic variables.

In the presence of a quantized vortex, the density of the system
drops to zero at its core,
whose size is characterized by a healing length $\xi$.
For very large condensates the healing length can be approximated by
$\xi=(8 \pi \rho_0 a)^{-1/2}$,
where $\rho_0$ is the density of the condensate before creating the
vortex. For a centered vortex line in the TF approximation
$\rho_0=\mu/g$, and the corresponding healing length $\xi_0$ is
\begin{equation}
\frac{\xi_0}{R_{\rm TF}}=\left(\frac{a_\perp}{R_{\rm TF}}\right)^2
\,\,\, ,
\label{healing}
\end{equation}
with $\xi_0 \ll a_\perp \ll R_{TF}$. In this approximation
a local healing length can be defined\cite{Lundh00} as
$\xi(r,z)=\xi_0/{\sqrt{1-(r/R_{\rm TF})^2-(z/Z_{\rm TF})^2}}$.
Note that the size of the core is larger for a vortex in the
low-density region.

\section{Superfluid velocity field}

The velocity field around a straight vortex line has an
analytical expression when the vortex is along the symmetry axis
\cite{fe01,Dalfovo96}. Approximate analytical expressions can be found
in the case of vortex lines off the symmetry axis in large and  very
elongated, quasi-two dimensional condensates
(see Ref.~\cite{gui2001} and references therein).
However, vortex lines generally bend in three-dimensional condensates
\cite{fe01,curva} which renders impracticable an analytical
treatment of the velocity field.

In the case of quantized vortex rings we proceed as Schwarz and
Jang have done in the case of helium II \cite{Schwarz} (see also
Ref. \cite{je98}). For a vortex ring characterized by the values of $(R,Z)$
already defined, circulation number $n= 1, 2, \ldots$ and quantum
circulation $k_0 = n\, h/m$, we write the vorticity in the line-source
approximation:
\begin{equation}
\mbox{\boldmath $\omega$} = k_0 \,\delta (r-R)\, \delta (z-Z) \,
\hat{\phi}     \,\,\, ,
\label{ringvor}
\end{equation}
where ($r,z,\phi$) are the cylindrical coordinates,
and $\hat{\phi}$ is the unit vector in the azimuthal direction
$\hat{\phi}=(-\sin\phi, \cos\phi, 0)$.
The superfluid velocity field that arises from this distributed vorticity
fulfills  $\mbox{\boldmath $\omega$} =\nabla \times \bf{v}$.
Hence, the velocity field around the vortex is irrotational except
on the vorticity line, where the density of the condensate is zero.
If ${\bf \nabla} \cdot {\bf v} = 0$ to a good approximation\cite{incom},
a velocity vector potential ${\bf A}({\bf r})$ can be introduced
such that ${\bf v} = {\bf \nabla} \times {\bf A}$.
If the vorticity is specified, ${\bf A}({\bf r})$ is determined by
the equation
\begin{equation}
\mbox{\boldmath $\omega$}= {\bf \nabla} \times ({\bf \nabla}
                 \times {\bf A}) \,\,\, ,
\label{eq1}
\end{equation}
whose integral solution is \cite{Schwarz}
${\bf A}({\bf r}) = A_0 (r,z) \hat{\phi}$, with
\begin{equation}
A_0(r,z) = \frac{k_0}{4 \pi} R
\int^{2 \pi}_{0} \frac{ \cos \phi\,' d\phi\,'}{\sqrt{r^2 + R^2 -
2 \,r R \,\cos \phi\,' + (z-Z)^2}}
\, \, .
\label{eq3}
\end{equation}
The radial and $z$-components of the velocity are obtained as
\begin{eqnarray}
v_r & = & - \frac{\partial A_0}{\partial z}
\nonumber
\\
& &
\label{eq4}
\\
v_z & = & \frac{1}{r} \frac{\partial}{\partial r} ( r A_0) \,\,,
\nonumber
\end{eqnarray}
whereas the azimuthal component of the velocity field around the
vortex ring Eq. (\ref{ringvor}) is zero.
We give in the Appendix the general expressions of $A_0$,
$v_r$ and $v_z$  written in terms of hypergeometric
functions \cite{Gr80}.

For a confined system, the existence of a boundary and the fact that
the density is inhomogeneous may have an effect on the actual velocity
field\cite{Ang01}. In our case, since the condensate extends up to
`infinite distances', there is no need to introduce any image vortex to
ensure that there is no particle flow across the boundary due to the
superfluid motion. Moreover, it has been shown\cite{Lundh00} that in the
TF limit the corrections to the velocity due to density inhomogeneities
can be safely neglected. Thus, we approximate the velocity field by
Eqs.~(\ref{eq3}) and (\ref{eq4}) (see Appendix), but have restricted
ourselves to study vortex ring configurations ($R,Z$) whose vorticity
line is inside the domain where the density is positive in the TF
approximation, i.e., only vortex rings $(R,Z)$ satisfying the condition
$(R/R_{\rm TF})^2+(Z/Z_{\rm TF})^2<1$ are considered.
We call the ($R,Z$) line defined by the condition
$(R/R_{\rm TF})^2+(Z/Z_{\rm TF})^2=1$ the TF boundary.
Once the the superfluid velocity field has been fixed, the density
profile of the condensate is obtained solving Eq.~(\ref{GP}).

\section{Results}

We present numerical results for large condensates
in the Thomas-Fermi limit hosting a quantized
vortex ring with circulation number equal to one
(nucleating vortex rings with $n>1$ is energetically less
favorable \cite{Donnelly,Winiecki99}).
For a singly quantized vortex ring
configuration $(R,Z)$ with vorticity given by Eq. (\ref{ringvor}) and
circulation number $n=1$,
we have computed the velocity field, Eq.~(\ref{eq4}), and
have obtained the density profile of the confined
condensate by solving Eq.~(\ref{GP}) in cilyndrical coordinates using
the imaginary time method\cite{Pi01}. Note that although we are in the
TF limit, we do solve the complete GP equation to obtain the density
profiles.

\begin{figure}
\centerline{\epsfig{figure=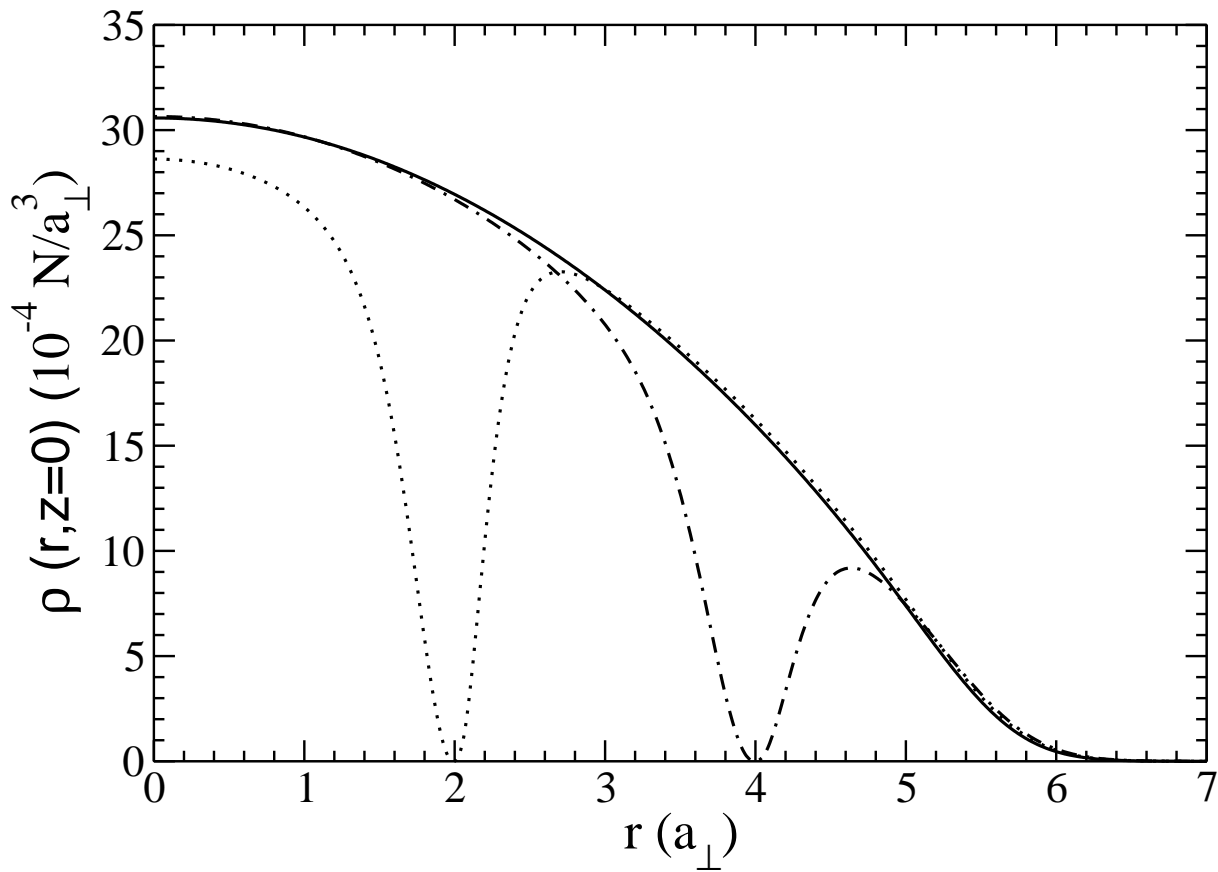,width=14.cm,angle=-90}}
\caption[]{
Density profile as a function of $r$ at $z=0$ for
a condensate in a spherical trap with the experimental parameters of
Ref.~\cite{an01} ($aN/a_\perp=440$) hosting a singly quantized vortex
ring at $Z=0$.
The dotted line corresponds to a vortex ring with $R=2 \, a_{\perp}$,
and the dot-dashed line to a vortex ring with $R=4 \, a_{\perp}$.
The solid line is the g.s. density profile in the absence
of vortices. The  densities have been normalized to one.
}
\label{fig1}
\end{figure}

Figure \ref{fig1} shows density profiles in the $z=0$
plane as function of $r$ for
the experimental parameters of Ref.~\cite{an01}, that is,
$N=3 \times 10^5$ atoms of $^{87}$Rb
(scattering length $a=5.82 \times 10^{-9}$~cm)
in a spherical trap with $\omega_\perp/2 \pi=7.8$ Hz.
We have plotted two configurations with a vortex ring located
in $Z=0$ having a radius $R=2 \, a_{\perp}$ (dotted line),
and $R=4 \, a_{\perp}$ (dot-dashed line), respectively.
The density profile of the condensate without vortex
is also shown (solid line).

The density is zero on the vorticity line, as
seen in the density profiles. As expected,
the size of the core is of the order of the healing length. Indeed,
for this condensate one has $R_{\rm TF} \simeq 5.8 \, a_{\perp}$ and
$\xi_0 \sim 0.17 \,a_{\perp}$.
Assuming that the vortex diameter is twice the local healing length,
and using the TF expressions we get as core diameters the values
$2 \times \xi(2,0)=0.37 \,a_{\perp}$, and
$2 \times \xi(4,0)=0.47 \,a_{\perp}$. One can see from Fig.~\ref{fig1}
that the core sizes are in agreement with these estimates.
When the radius of the vortex ring increases, the
core lies at a lower density region and therefore its
size increases.

\begin{figure}
\centerline{\epsfig{figure=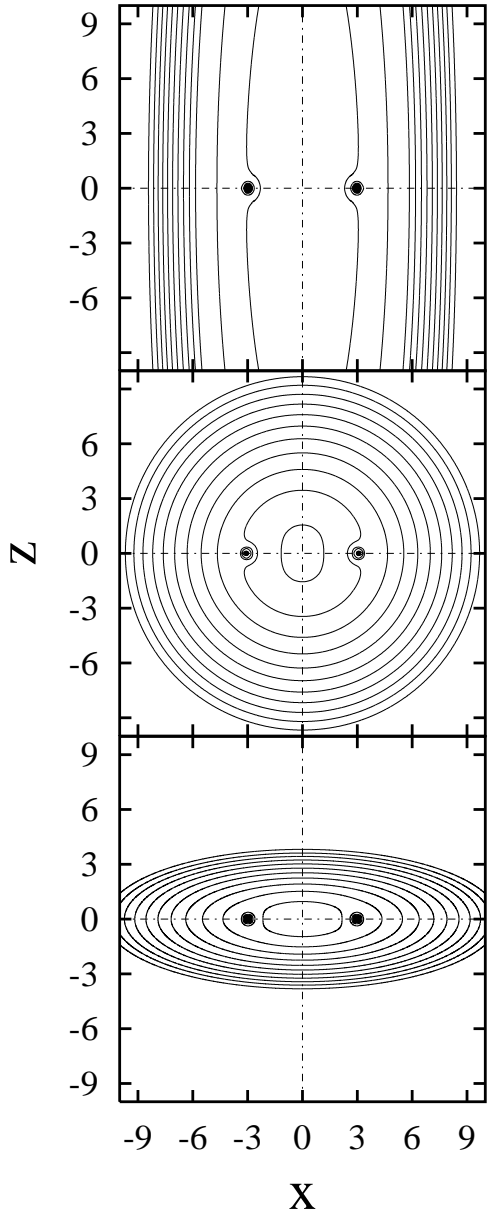,width=12.5cm}}
\vspace*{1.cm}
\caption[]{
Vortex ring state equidensity lines, in arbitrary units, plotted in
the $y=0$ plane (axes are in units of $a_\perp$).
The ring is at ($R=3.1\,a_\perp, Z=0$) in
a condensate with $N=10^6$ atoms of $^{87}$Rb, confined in an axially
symmetric trap with axial frequency $\omega_z/2 \pi =220$ Hz and
different trap geometries.
The top panel corresponds to a cigar-shaped trap
($\lambda=0.2$, $R_{\rm TF}=8.7 \, a_\perp, \,Z_{\rm TF}=43.3 \,
a_\perp$),
the middle panel to a spherically symmetric trap ($\lambda=1$,
$R_{\rm TF}=10.2 \, a_\perp$), and the bottom panel  to a disk-shaped
trap ($\lambda=\sqrt 8, \,R_{\rm TF}=45.5 \, a_\perp, \,
Z_{\rm TF}=11.3 \, a_\perp$).
}
\label{fig2}
\end{figure}

To study the  vortex ring energetics we have chosen a larger condensate
made of $N= 10^6$ atoms of $^{87}$Rb confined in an axially symmetric
trap with axial frequency $\omega_{z}/2 \pi = 220$ Hz and
three different geometries, namely  spherical, disk-shaped and
cigar-shaped, with $\lambda=1, \sqrt 8$ and $0.2$, respectively.
For this condensate, we show in Fig. \ref{fig2} several equidensity
lines (arbitrary values) in the $y=0$ plane.
The top panel corresponds to the cigar-shaped trap
($\lambda=0.2$, $R_{\rm TF}=8.7 \, a_\perp$, $Z_{\rm TF}=43.3 \,
a_\perp$), the middle panel to the spherically symmetric trap
($\lambda=1$,
$R_{\rm TF}=Z_{\rm TF}=10.2 \, a_\perp$), and the
bottom panel to the disk-shaped trap ($\lambda=\sqrt 8$,
$R_{\rm TF}=45.5 \, a_\perp$, $Z_{\rm TF}=11.3 \, a_\perp$).
The $x$ and $z$ axes are in units of $a_\perp$. For all geometries, the
vortex ring configuration is ($R =3.1 \, a_\perp, Z=0)$. The
intersection between the vorticity line and the $y=0$ plane appears
as two dark dots, indicating the steep density depression
around the vortex core.
The presence of the vorticity causes a drastic distortion of the
density with respect to the g.s. profile. Yet, for large condensates it is
a rather local effect, as can be seen from Fig. \ref{fig2}, and also
from Fig. \ref{fig1} for the smaller condensate.

\begin{figure}
\centerline{\epsfig{figure=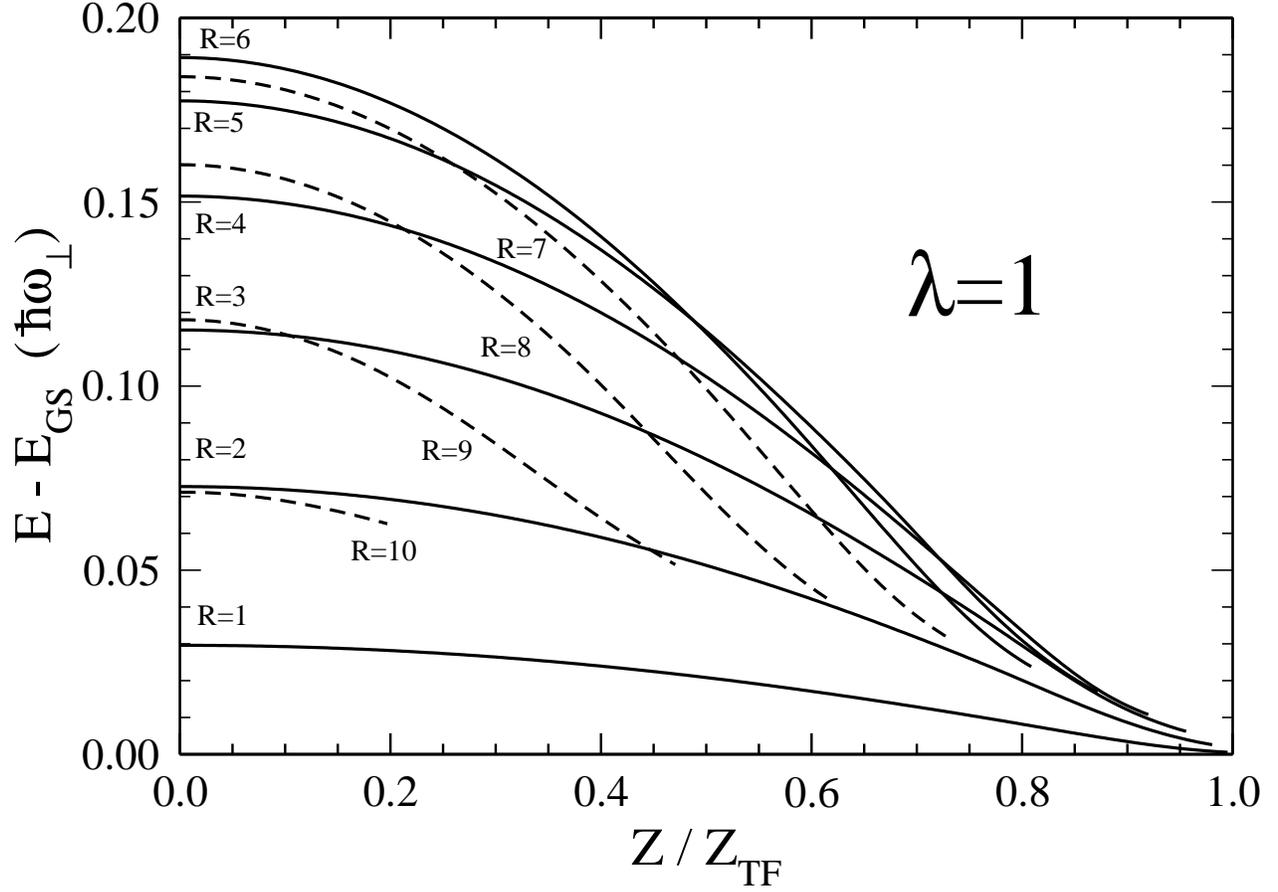,width=14.cm,angle=-90}}
\caption[]{ Nucleation energy of a vortex ring $(R,Z)$
as a function of $Z$ for different values of the radius.
The labels of the curves are in units of $a_\perp$.
The condensate has $N=10^6$ atoms of $^{87}$Rb
and it is confined in an spherically symmetric trap
with frequency  $\omega_z/2 \pi =220$ Hz. The TF radius is
$R_{\rm TF}=Z_{\rm TF}=10.2 \, a_\perp$.
}
\label{fig3}
\end{figure}

The nucleation of a vortex has an energy cost,
since the energy of a condensate with a vortex is always
larger than the  energy of the condensate without it, $E_{\rm GS}$.
The vortex nucleation energy, $E-E_{\rm GS}$, corresponding to
the condensate of the spherical trap in Fig. \ref{fig2} is
plotted in Fig. \ref{fig3} as a function of $Z$ for different values
of $R$. Fixed $R$, the nucleation energy is maximum when the ring is
in the $z=0$ plane, and it decreases as $Z$ increases, since displacing
the ring outwards implies that the superfluid flow affects less atoms
in the condensate.
Note that to have the vorticity line within the TF boundary,
the $R$-curves end at different values of $Z/Z_{\rm TF}$
given by $\sqrt{1-(R/R_{\rm TF})^2}$.
The nucleation energy does not vanish when the ring is located
on the TF boundary as there still exists a superfluid
velocity field which produces some effect. Since we treat
the superfluid velocity as an external field, the nucleation
energy of very superficial vortex configurations may be
somewhat overestimated. This drawback is apparent beyond the TF boundary,
and shows up as a long-tailed nucleation energy as a function of either
$Z/Z_{TF}$ or $R/R_{TF}$, although it goes eventually to zero.

\begin{figure}
\centerline{\epsfig{figure=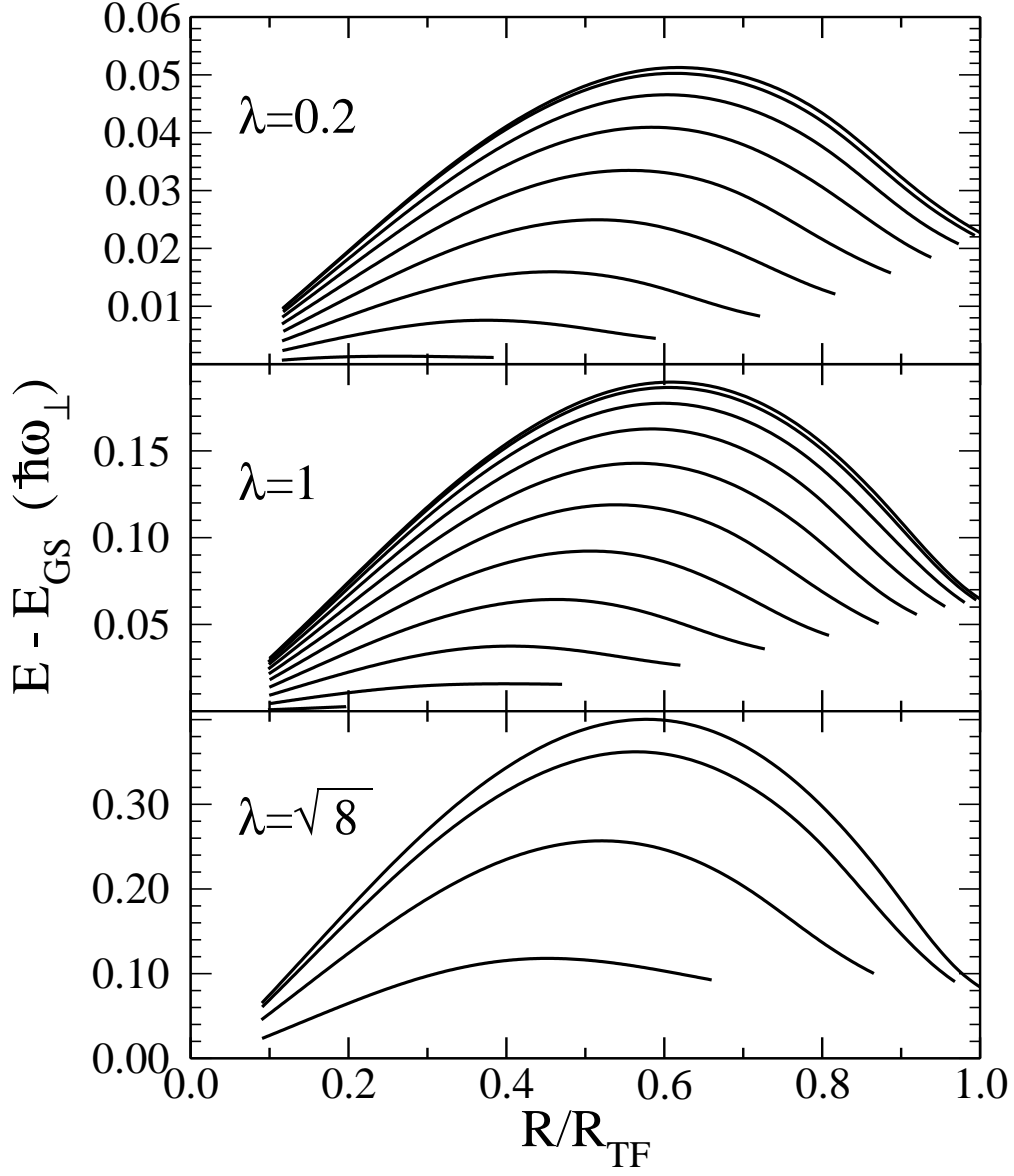,width=14.cm}}
\caption[]{ Nucleation energy of a vortex ring $(R,Z)$
as a function of $R$ for different values of $Z$.
The condensate and trap geometries are these of Fig. \ref{fig2}.
The top panel corresponds to the cigar-shaped trap,
the middle panel to the spherically symmetric trap,
and the bottom panel to the disk-shaped trap. From top to
bottom, and in units of $a_\perp$, the curves correspond
to $Z=0$ to 40 in $Z$-steps of 5 (top panel);
$Z=0$ to 10 in $Z$-steps of 1 (middle panel);
$Z=0, 1, 2$, and 3 (bottom panel).
}
\label{fig4}
\end{figure}

The nucleation energy is plotted in Fig. \ref{fig4} as a function
of $R$ for different $Z$ values. The smaller ring we have considered
has $R/R_{\rm TF} \sim 0.1$. The middle panel
corresponds to the same spherical condensate as in Fig.~\ref{fig3}.
Besides, the top panel corresponds to the cigar-shaped trap,
and the bottom panel to the disk-shaped trap.
The top curve in all panels corresponds to vortex rings with $Z=0$, and
for a given trap geometry (fixed panel), the higher the curve, the
lower the $Z$. As in Fig. \ref{fig3}, we have only described
vortex ring
configurations inside the TF boundary. For this reason, fixed a
$Z$ value the corresponding nucleation energy curve stops at the
value $R/R_{\rm TF}=\sqrt{1-(Z/Z_{\rm TF})^2}$.

Figure \ref{fig4} shows that for each asymmetry parameter
$\lambda$, given a
$Z$ there exists a vortex configuration corresponding to a
certain radius $R$ that maximizes the nucleation energy. This radius
increases as $Z$ decreases. In the ($R,Z$) two-dimensional configuration
space, the nucleation energy $E(R,Z)-E_{\rm GS}$
has only one maximum at $(R_{\rm eq}/R_{\rm TF}\simeq 0.6,Z=0)$.
It is in agreement with the results obtained in Ref. \cite{ja992} for a
spherically symmetric trap using a simplified model of a vortex ring.

This maximum corresponds to the only stationary, although unstable,
vortex ring configuration
(equivalent to a vortex line nucleated along the $z$-axis),
whose location is almost independent of $\lambda$
but its value depends on it, decreasing as $\lambda$ does.
This means that the energy to nucleate a vortex ring characterized
by the values of $(R,Z)$ is lower the more elongated the trap is.

For each asymmetry parameter, the `iso-nucleation energy' lines in the
($R,Z$) configuration  space corresponding to a value not too far from
the maximum value are closed curves around the maximum, and the allowed
ring radii are in the range defined by the intersection of a
straight line  representing the given nucleation energy and the $Z=0$
curve in Fig. \ref{fig4}. If we relax the condition that the
vorticity line must be inside the TF boundary, this should be always
like this\cite{ja992}. Otherwise, the isoenergy lines eventually do
not close. Yet, the allowed ring radii are determined in a similar way.

We have considered the vortex ring as if it were a static object.
However, in a trapped condensate a vortex ring will
move with a velocity that results from the interplay
between the effect caused by the inhomogeneity of the condensate and
self-induced effects arising from its own local curvature
\cite{fe01,Donnelly,ja992}.
A qualitative analysis of the dynamics of coaxial  vortex rings can be
carried out from Figs.~\ref{fig3} and \ref{fig4}, as proposed in
Ref.~\cite{ja992}, and tested in a real time dynamics for
two-component condensates\cite{ru01,ja992}.
It can be seen that a coaxial vortex ring oscillates inside the
condensate, moving towards the surface and instead of being
annihilated at the boundary, the ring generates a background flow in
the rest of the cloud that draws it back towards the center.
The ring dynamics in a trapped condensate results in an oscillation
along the symmetry axis of the trap. The radius of the vortex ring
grows or shrinks depending on its $Z$-position, in such a way
that the ring motion almost follows
a trajectory with constant energy in the $(R,Z)$
configuration space.

Finally,  we have used a completely different method, numerically more
involved, to generate quantized vortex rings. It does not
make any assumption on the superfluid velocity field, nor on the radius
or position of the vortex ring either, and can be used to test the
results we have obtained based on the validity of the
approximation ${\bf \nabla} \cdot {\bf v}=0$.

We have proceeded by analogy with the case of a vortex line, and have
added a constraint to the energy functional that forces
to form a toroidal-like hole in the condensate with a flux of atoms around
its core with zero azimuthal velocity component.
We have used as constraining operator the angular momentum
about the $\hat{\phi}$ axis, i.e., $\hat{\phi} \cdot {\bf L} \equiv
L_\phi=i \hbar (r \partial_z - z \partial_r)$\cite{hermit}.

Introducing the functional
\begin{equation}
E[\,\Psi\,] = \int d{\bf r} \left[\frac{\hbar^2}{2 m} |\nabla \Psi|^2 +
 V_{\rm ext}({\bf r}) |\Psi|^2 + \frac{g}{2} |\Psi|^4 \right] +
   \Omega_\phi \int d \, {\bf r} \, \Psi^*
       L_\phi \Psi  \,\,\, ,
\label{newfunct}
\end{equation}
where the Lagrange multiplier $\Omega_\phi$ can be understood
as the local angular velocity about the azimuthal direction, we
have obtained the associated GP equation.
For a given $\Omega_\phi$, we have solved the
partial differential equations obeyed by
the real and imaginary part of the condensate wave function, which
are coupled by the presence of the constraint.
These equations have been discretized on a $(r,z)$ mesh using
seven-point formulas to represent the differential operators, and have
been solved using again the imaginary time method\cite{Pi01}.

Taking as an example $\Omega_\phi = 0.2 \,\omega_\perp$,
we have obtained a coaxial vortex ring configuration with its
vorticity located on the $z=0$ plane whose
equidensity lines cannot be distinguished from these represented in
the middle panel of Fig. \ref{fig2} at the scale of this figure (whose
parameters were chosen before to allow us to make this statement).

\begin{figure}
\centerline{\epsfig{figure=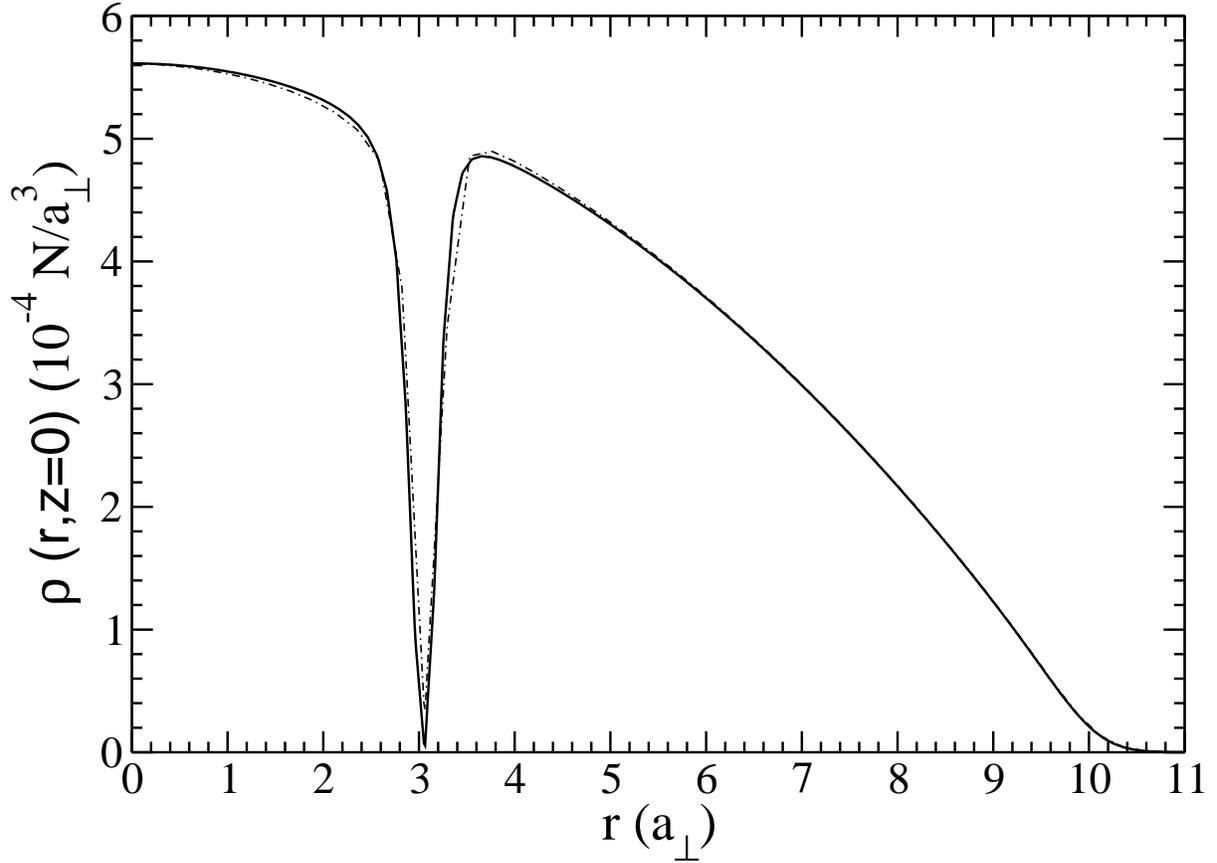,width=14.cm,angle=-90}}
\caption[]{
Density profile as function of $r$ at $z=0$ for the $N=10^6$
condensate in a spherically symmetric trap
with a singly quantized vortex ring.
The solid line corresponds to the density profile that minimizes
the functional Eq. (\ref{newfunct}) with constraint
$\Omega_\phi=0.2 \,\omega_\perp$.
The dashed line is the density obtained  imposing a
ring vorticity at ($R= 3.1\, a_{\perp}, Z=0$).
The densities have been normalized to one.
}
\label{fig5}
\end{figure}

Figure \ref{fig5} presents a comparison between
the two methods. It shows the particle density profile as a function of
$r$ at $z=0$ for the spherical condensate of Fig.~\ref{fig3}.
The solid line is the configuration
that minimizes Eq. (\ref{newfunct}). The density goes to zero in
$R=3.1 \, a_\perp$, and the vortex ring
has an energy equal to $37.164 \, \hbar \omega_\perp$.
The dashed line is the configuration obtained by the method
of Sect.~2 with the ring vorticity
Eq. (\ref{ringvor}) placed in $(R=3.1 \, a_\perp, Z=0$) which has
an energy of $37.167 \, \hbar \omega_\perp$.
Thus, the energy and density profile of both
configurations are almost identical, and the
velocity fields are also in good agreement.

\begin{figure}[t,H]
\centerline{\epsfig{figure=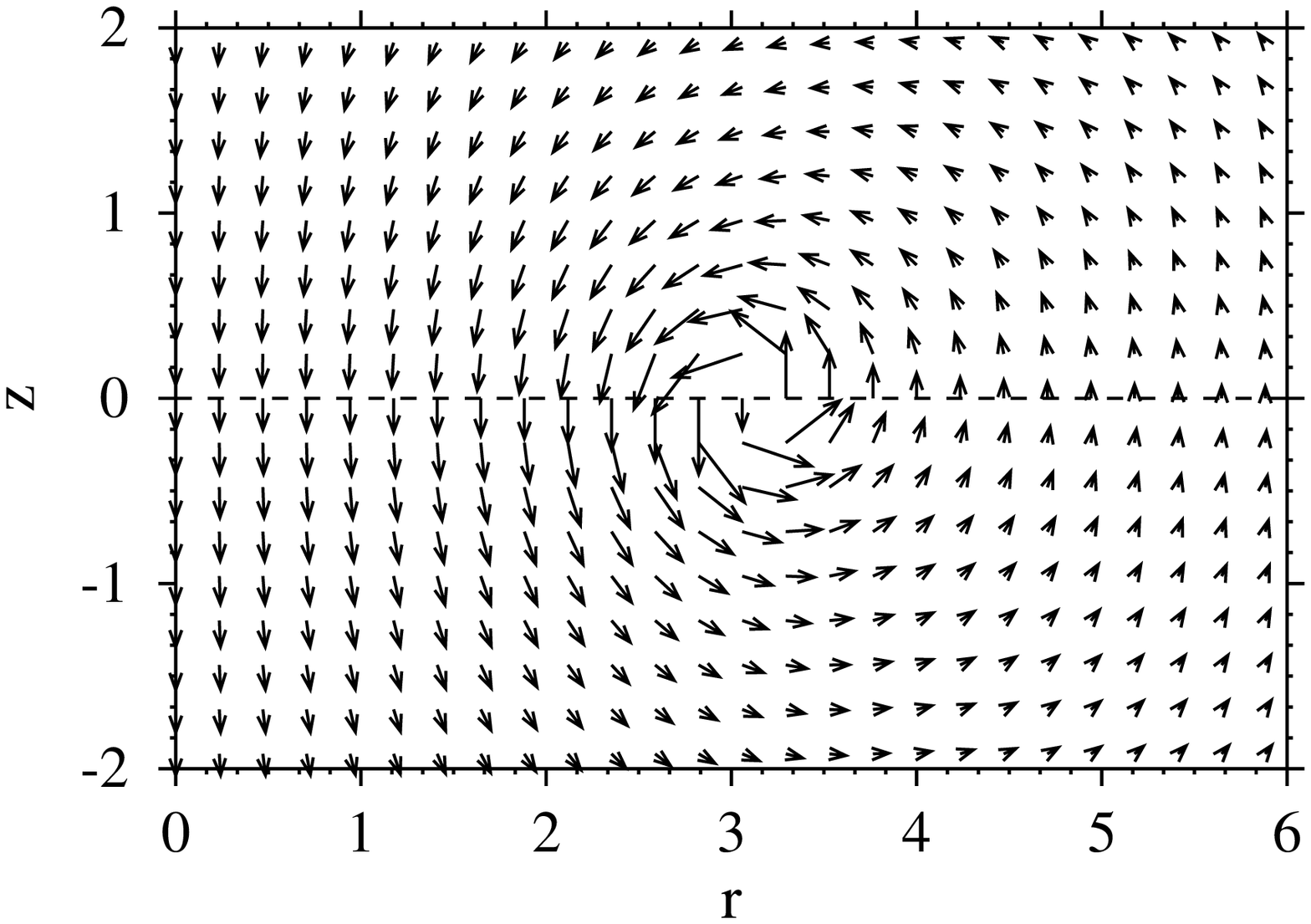}}
\vspace*{1.cm}
\caption[]{
Current field around the vortex core corresponding to the vortex ring
state calculated taking $\Omega_\phi=0.2\, \omega_\perp$ in Eq.
(\ref{newfunct}).
The current is in arbitrary units and the axes are in units of
$a_{\perp}$.
}
\label{fig6}
\end{figure}

We have also checked that minimizing the functional Eq. (\ref{newfunct})
yields quantized ring vortices with $n=1$. To this end, we have
proceeded to a direct numerical integration of the circulation
$\oint {\bf v} \cdot d{\bf l}$ using several arbitrary closed
paths in the ($r,z$) `plane'. We have found that the circulation
for paths enclosing the vortex core is $h/m$ with a good
accuracy, and is zero otherwise. The velocity field has been
calculated from the wave function recalling that the current field
is ${\bf j}({\bf r})= \rho\, {\bf v}=(\hbar/2 m i)[\Psi^* \mbox{\boldmath
$\nabla$} \Psi -  (\mbox{\boldmath $\nabla$} \Psi^*) \Psi]$.
The current field ${\bf j}({\bf r})$ in the $y=0$ plane is plotted in
arbitrary units in Fig. \ref{fig6}. The axes are in units of $a_\perp$.
It can be clearly distinguished in this figure the position of
the vortex ring core at $Z=0$ and $R \simeq 3.1 \, a_{\perp}$.

\section{summary}

We have developed a method to study vortex rings hosted in
confined condensates in the TF limit. It provides analytical
expressions for the superfluid velocity field that satisfy the
irrotational condition and the quantization of the circulation of the
superfluid flow.
The velocity around the vortex core is introduced as an external
field in the Gross-Pitaevskii energy functional, which is minimized
to obtain the wave function of the condensate.

Using this method, we have computed  the density profile and
nucleation energy of vortex rings in one-component condensates.
We have found that the size of the vortex ring core is of the order of
the healing length as in the case of vortex lines. The presence of the
vortex causes a sizeable distortion of the density, as
atoms are pushed off the core, but the effect is rather local.

The analysis of the dependence of the nucleation energy
on the position and
radius of the ring allows one to conclude that a Hamiltonian
dynamics will lead to oscillations  of the vortex ring along the
symmetry axis of the condensate, increasing and decreasing its radius
in accordance with previous predictions. A dissipative dynamics would
cause the vortex to decay, for example disappearing
across the border of the condensate, as occurs for vortex
lines\cite{Fed99}.

The method provides a handleable way to generate vortex rings
in confined condensates that can be used as a starting point to study
the dynamics. A detailed calculation of the vortex ring dynamics
in a confined condensate, either Hamiltonian or dissipative,
is beyond the scope of the present work and will be
developed elsewhere.

\section*{acknowledgements}
This work has been performed under Grants No. PB98-1247 from DGESIC,
Spain, and No. 2000SGR-00024 from Generalitat de Catalunya.
D.M.J. acknowledges  the CONICET (Argentina) and the Generalitat de
Catalunya ACI program for financial support.

\appendix
\section*{}

In this Appendix we give the expressions and method we have used
to obtain the velocity field. From
Eqs.~(\ref{eq3}-\ref{eq4}) one has:
\begin{equation}
v_r(r,z) = \frac{k_0\, R\,(z-Z)}{\pi \left[(r+R)^2+(z-Z)^2\right]^{3/2}}
\int^{\pi/2}_0 \frac{2 \,\cos^2\phi - 1}{(1 - s \,\,\cos^2\phi)^{3/2}}
\,d\phi      \,\,\, ,
\label{a1}
\end{equation}
where we have defined $s \equiv 4 r R/ \left[(r+R)^2+(z-Z)^2\right]$.
The integral Eq.~(\ref{a1}) is written in terms of hypergeometric
functions \cite{Gr80} yielding:
\begin{equation}
v_r(r,z) = \frac{k_0\, R\,(z-Z)}{2 \left[(r+R)^2+(z-Z)^2\right]^{3/2}}
\left[ F(3/2, 3/2; 2; s) - F(1/2, 3/2; 1; s)\right]     \,\, .
\label{a2}
\end{equation}
Similarly,
\begin{eqnarray}
v_z(r,z) & = & \frac{k_0\, R}{2\, r \left[(r+R)^2+(z-Z)^2\right]^{1/2}}
\left[ F(3/2, 1/2; 2; s) - F(1/2, 1/2; 1; s)\right]
\nonumber
\\
& + & \frac{k_0\, R}{2 \left[(r+R)^2+(z-Z)^2\right]^{3/2}}
\left[ (r+R)\, F(1/2, 3/2; 1; s) - (r+2R)\,F(3/2, 3/2; 2; s)\right.
\nonumber
\\
& + & \frac{3R}{2}\, F(5/2, 3/2; 3; s) ]
\label{a3}
\end{eqnarray}
and
\begin{equation}
A_0(r,z)   =   \frac{k_0\, R}{2 \left[(r+R)^2+(z-Z)^2\right]^{1/2}}
\left[ F(3/2, 1/2; 2; s) - F(1/2, 1/2; 1; s)\right]  \,\, .
\label{a4}
\end{equation}
The hypergeometric functions entering Eqs.~(\ref{a2}-\ref{a4}) can be
written in terms of the complete elliptic integrals {\bf E} and
{\bf K} \cite{Gr80} as follows:
\begin{equation}
F(1/2,1/2;1;s)= \frac{2}{\pi} {\bf K}(\sqrt{s})
\label{a5}
\end{equation}
\begin{equation}
F(1/2,3/2;1;s)  =  F(3/2,1/2;1;s)
 =   \frac{2}{\pi} \frac{{\bf E}(\sqrt{s})}{1-s}
\label{a6}
\end{equation}
\begin{equation}
F(3/2,1/2;2;s)   =   F(1/2,3/2;2;s)
  =    \frac{4}{\pi s} [{\bf K}(\sqrt{s})- {\bf E}(\sqrt{s}) ]
\label{a7}
\end{equation}
\begin{equation}
F(3/2,3/2;2;s)  =  \frac{4}{\pi s} \left[
\frac{{\bf E}(\sqrt{s})}{1-s} -{\bf K}(\sqrt{s}) \right]
\label{a8}
\end{equation}
\begin{equation}
F(5/2,3/2;3;s)  =  \frac{16}{3 \pi s^2} \left[
\frac{2-s}{1-s}{\bf E}(\sqrt{s}) -2 {\bf K}(\sqrt{s}) \right] \,\, .
\label{a9}
\end{equation}
To evaluate {\bf E} and {\bf K} we have used  polynomial approximations
\cite{AS}.


%
\pagebreak


\begin{references}

\bibitem{ma99}
M. R. Matthews, B. P. Anderson, P. C. Haljan, D. S. Hall, C.E.
Wieman, and E. A. Cornell, Phys. Rev. Lett. {\bf 83}, 2498 (1999).

\bibitem{an00}
B. P. Anderson, P. C. Haljan, C. E. Wieman, and E. A. Cornell,
 Phys. Rev. Lett. {\bf 85}, 2857 (2000).

\bibitem{da} K. W. Madison, F. Chevy, W. Wohlleben, and
J. Dalibard, Phys. Rev. Lett. {\bf 84}, 806 (2000).

\bibitem{ke} J. R. Abo-Shaeer, C. Raman, J. M. Vogels, and
W. Ketterle. Science {\bf 292}, 476 (2001).

\bibitem{fe01} A. L. Fetter and A. A. Svidzinsky,
J. Phys.: Condens. Matter {\bf 13}, R135 (2001).

\bibitem{Bau95}
G. H. Bauer, R. J. Donnelly, and W. F. Vinen, J. Low Temp.
Phys. {\bf 98}, 47 (1995).

\bibitem{Dal00}
F. Dalfovo, R. Mayol, M. Pi, and M. Barranco,
Phys. Rev. Lett. {\bf 85}, 1028 (2000).

\bibitem{May01}
R. Mayol, M. Pi, M. Barranco, and F. Dalfovo,
Phys. Rev. Lett. {\bf 87}, 145301 (2001).

\bibitem{Donnelly} R. J. Donnelly, {\it Quantized Vortices in Helium
II}, (Cambridge University Press, Cambridge, 1991).


\bibitem{ja991}
B. Jackson, J. F. McCann, and C. S. Adams, Phys. Rev. A
 {\bf 60}, 4882 (1999).

\bibitem{fe00}
D. L. Feder, M. S. Pindzola, L. A. Collins, B. I. Schneider,
and C. W. Clark, Phys. Rev. A {\bf 62}, 053606 (2000).

\bibitem{ru01}
J. Ruostekoski and J. R. Anglin, Phys. Rev. Lett.
 {\bf 86}, 3934 (2001).

\bibitem{an01}
B. P. Anderson, P. C. Haljan, C. A. Regal, D. L. Feder, L. A. Collins,
C. W. Clark, and E. A. Cornell, Phys. Rev. Lett. {\bf 86}, 2926
(2001).

\bibitem{da99} F. Dalfovo, S. Giorgini, L. Pitaevskii, and S. Stringari,
Rev. Mod. Phys. {\bf 71}, 463 (1999).

\bibitem{Lundh00} E. Lundh and P. Ao,
Phys. Rev. A {\bf 61}, 063612 (2000).

\bibitem{Dalfovo96} F. Dalfovo and S. Stringari,
Phys. Rev. A {\bf 53}, 2477 (1996).

\bibitem{gui2001} M. Guilleumas and R. Graham,
Phys. Rev. A {\bf 64}, 033607 (2001).

\bibitem{curva} J. J. Garc\'{\i}a-Ripoll and V. M. P\'erez-Garc\'{\i}a,
Phys. Rev. A {\bf 63}, 041603(R) (2001).

\bibitem{Schwarz} K. K. Schwarz and P. S. Jang,
Phys. Rev. A {\bf 8}, 3199 (1973).

\bibitem{je98} D. M. Jezek, M. Pi, M. Barranco, R. J. Lombard,
and M. Guilleumas, J. Low Temp. Phys. {\bf 112}, 303 (1998).

\bibitem{incom} Note that since the density is not uniform,
the assumption ${\bf \nabla} \cdot {\bf v} \approx 0$
does not imply that the system is incompressible.

\bibitem{Gr80} I. S. Gradshteyn and I. M. Ryzhik, {\it Table of
Integrals, Series and Products} (Academic Press, New York, 1980).

\bibitem{Ang01} J. R. Anglin, e-print cond-mat/0110389.

\bibitem{Winiecki99} T. Winiecki, J. F. McCann, and C. S. Adams,
Europhys. Lett., {\bf 48}, 476 (1999).

\bibitem{Pi01} M. Pi, A. Emperador, M. Barranco, and F. Garcias,
Phys. Rev. B {\bf 63}, 115316 (2001).

\bibitem{ja992}
B. Jackson, J. F. McCann, and C. S. Adams, Phys. Rev. A
 {\bf 61}, 013604 (1999).

\bibitem{hermit}
Actually, we have used the hermitian form
$L_\phi=(\hat{\phi} \cdot {\bf L}+ {\bf L} \cdot \hat{\phi})/2$.

\bibitem{Fed99} P. O. Fedichev and G. V. Shlyapnikov
Phys. Rev. A {\bf 60}, R1779 (1999).

\bibitem{AS} M. Abramowitz and I. Stegun, {\it Handbook of
Mathematical Functions} (Dover Pu., New York, 1970).

\end{references}
\end{document}